\documentclass[]{spie}  

\usepackage{amsmath,amsfonts,amssymb}
\usepackage{graphicx}
\usepackage[colorlinks=true, allcolors=blue]{hyperref}

\title{Deep observations with an ELT in the Global Multi Conjugated Adaptive Optics perspective}

\author[a,b]{Elisa Portaluri}
\author[a,b]{Valentina Viotto}
\author[a,b]{Roberto Ragazzoni}
\author[a,b]{Carmelo Arcidiacono}
\author[a,b]{Maria Bergomi}
\author[a,b]{Marco Dima}
\author[a,b]{Davide Greggio}
\author[a,b]{Jacopo Farinato}
\author[a,b]{Demetrio Magrin}

\affil[a]{INAF - Osservatorio Astronomico di Padova - vicolo dell'Osservatorio, 35122, Padova Italy}
\affil[b]{ADONI - Laboratorio Nazionale di Ottica Adattiva, Italy}

\authorinfo{Further author information: (Send correspondence to EP)\\EP: E-mail: elisa.portaluri@inaf.it}

\begin{document} 
\maketitle

\begin{abstract}
Deep observations of the Universe, usually as a part of sky surveys, are one of the symbols of the modern astronomy because they can allow big collaborations, exploiting multiple facilities and shared knowledge. The new generation of extremely large telescopes will play a key role because of their angular resolution and their capability in collecting the light of faint sources. Our simulations combine technical, tomographic and observational information, and benefit of the Global-Multi Conjugate Adaptive Optics (GMCAO) approach, a well demonstrated method that exploits only natural guide stars to correct the scientific field of view from the atmospheric turbulence. By simulating K-band observations of 6000 high redshift galaxies in the Chandra Deep Field South area, we have shown how an ELT can carry out photometric surveys successfully, recovering morphological and structural parameters. We present here a wide statistics of the expected performance of a GMCAO-equipped ELT in 22 well-known surveys in terms of SR. 
\end{abstract}

\keywords{Instrumentation: adaptive optics - Surveys - Methods: statistical}

\section{INTRODUCTION}
\label{sec:intro}  

Sky surveys are a key instrument to study the deep Universe and understand how galaxies form and evolve. 
We investigated how the new generation of extremely large telescopes will perform in this field of research: they will play a key role because of their angular resolution and their capability in collecting the light of faint sources.

\begin{figure} [hb]
\begin{center}
\begin{tabular}{c} 
\includegraphics[height=6cm]{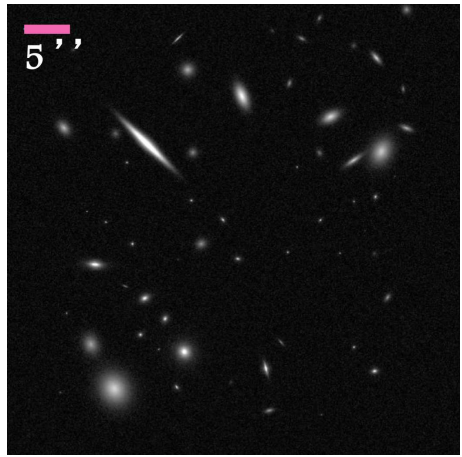}
\end{tabular}
\end{center}
\caption[example] 
{\label{fig:chandra} Mock K-band image of one portion of the Chandra-GMCAO deep field. The FoV is 50x50 arcsec$^2$. The magenta dash is 5 arcsec wide. 60 high-redshift galaxies are present. The figure is taken from the work of [\citenum{Portaluri2017}].}
\end{figure} 

However a big limitation of such kind of studies is the fact that the observations are taken in a dark area to enhance the light of distant and/or small objects. Such dark regions of the sky can be a test-bench for a number of Adaptive Optics techniques because of the star-poorness, mainly because a certain number of natural references is required even if the facility foresees the use of lasers.
GMCAO, avoiding the use of LGSs, has the advantage to be free from the associated issues ([\citenum{Fried1995},\citenum{Pfrommer2009},\citenum{Diolaiti2012}]).Starting from a feasibility study committed by ESO, we investigate the performance of the Global MCAO technique.

The method was introduced by [\citenum{Ragazzoni2010}] and followed by several studies focused on the technical specifications and evaluations ([\citenum{Viotto2015},\citenum{Viotto2016},\citenum{Viotto2018}]) and on the scientific performance ([\citenum{Portaluri2015},\citenum{Portaluri2016},\citenum{Portaluri2018},\citenum{Portaluri2018b}]). The method can be used in the next generation AO facilities like the European Extremely Large Telescope [\citenum{Gilmozzi2007}], the Giant Magellan Telescope [\citenum{Johns2008}], and the Thirty Meter Telescope [\citenum{Szeto2008}]. 

In particular, by simulating K-band observations of 6000 high redshift galaxies in the Chandra Deep Field South region (a portion of the mock image is shown in Figure~\ref{fig:chandra}), we have demonstrated how a GMCAO-equipped ELT can carry out photometric surveys successfully [\citenum{Portaluri2017}].
Here we extended the analysis using 22 well-known deep field surveys to get a more robust statistics on the results, showing how GMCAO can correct homogeneously star-poor regions, otherwise impracticable without the use of laser guide stars.

\begin{table}[hb]
\caption{List of the simulated GMCAO surveys campaign. Column~1: Name of the GMCAO field. Column~2: Right Ascension and Declination of the central pointing used by GIUSTO.} 
\label{surveys}
\begin{center}       
\begin{tabular}{|l|c|} 
\hline
\rule[-1ex]{0pt}{3.5ex}GMCAO \#1  &  53.12,-27.81   \\    
\hline
\rule[-1ex]{0pt}{3.5ex}GMCAO \#2  &  150.12, 2.21   \\    
\hline
\rule[-1ex]{0pt}{3.5ex}GMCAO \#3  &  338.25, -60.55 \\     
\hline
\rule[-1ex]{0pt}{3.5ex}GMCAO \#4  &  189.20, 62.22  \\     
\hline
\rule[-1ex]{0pt}{3.5ex}GMCAO \#5  &  34.41, -5.20   \\                       
\hline
\rule[-1ex]{0pt}{3.5ex}GMCAO \#6  &  214.25, 52.00  \\     
\hline
\rule[-1ex]{0pt}{3.5ex}GMCAO \#7  &  36.50, -4.49   \\                                            
\hline
\rule[-1ex]{0pt}{3.5ex}GMCAO \#8  &  333.88, -17.73 \\    
\hline
\rule[-1ex]{0pt}{3.5ex}GMCAO \#9  &  201.16, 27.49  \\	 
\hline
\rule[-1ex]{0pt}{3.5ex}GMCAO \#10 &  13.35, 12.57   \\    
\hline
\rule[-1ex]{0pt}{3.5ex}GMCAO \#11 &  139.50, 30.00  \\    
\hline
\rule[-1ex]{0pt}{3.5ex}GMCAO \#12 &  80.00, -49.00  \\    
\hline
\rule[-1ex]{0pt}{3.5ex}GMCAO \#13 &  163.00, -5.00  \\    
\hline
\rule[-1ex]{0pt}{3.5ex}GMCAO \#14 &  208.75, -10.00 \\    
\hline
\rule[-1ex]{0pt}{3.5ex}GMCAO \#15 &  32.50, -4.50   \\    
\hline
\rule[-1ex]{0pt}{3.5ex}GMCAO \#16 &  218.00, 34.28  \\    
\hline
\rule[-1ex]{0pt}{3.5ex}GMCAO \#17 &  217.5, 34.50   \\    
\hline
\rule[-1ex]{0pt}{3.5ex}GMCAO \#18 &  31.87, -4.74   \\	     			     
\hline
\rule[-1ex]{0pt}{3.5ex}GMCAO \#19 &  36.25, -4.50   \\    
\hline
\rule[-1ex]{0pt}{3.5ex}GMCAO \#20 &  164.25, 57.67  \\     
\hline
\rule[-1ex]{0pt}{3.5ex}GMCAO \#21 &  242.50, 54.00  \\ 
\hline
\rule[-1ex]{0pt}{3.5ex}GMCAO \#22 &  334.25, 0.33   \\
\hline
\end{tabular}
\end{center}
\end{table} 

\section{The method}
To select our sample, we overviewed the most famous and studied deep field surveys [\citenum{Portaluri2018b}] that, as said, having the star-poorness as a key and mandatory condition, are good test-benches for the GMCAO technique.
We built 22 (500x500 arcsec$^2$) mock fields, listed Table~\ref{surveys}, composed by 100 sub-fields of 50x50 arcsec$^2$ each to match the expected characteristics of ELT.

On each sub-field we run GIUSTO (GMCAO Interactive Data Language Unreleased Simulation TOol), an IDL tomographic simulation tool that, giving a number of technical and astrophysical input parameters, calculates the performance of a given telescope that benefits of GMCAO in terms of SR over a given FoV. A detailed description of the simulator can be found in [\citenum{Viotto2015}], together with some tests we performed to evaluate the best parameter we should adopted for the study.
A list of the main variables we determined are shown in the Table~\ref{param}.

\begin{figure} [h]
\begin{center}
\begin{tabular}{c} 
\includegraphics[height=6.5cm]{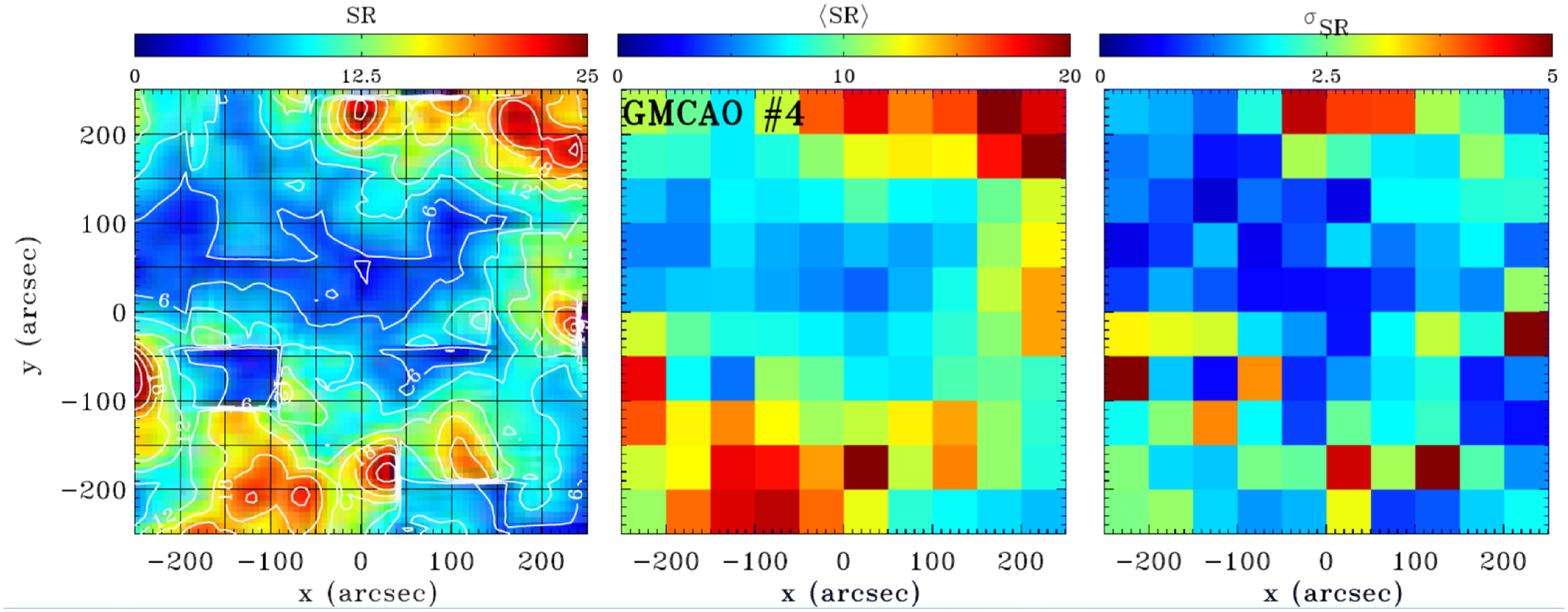}\\
\includegraphics[height=6.5cm]{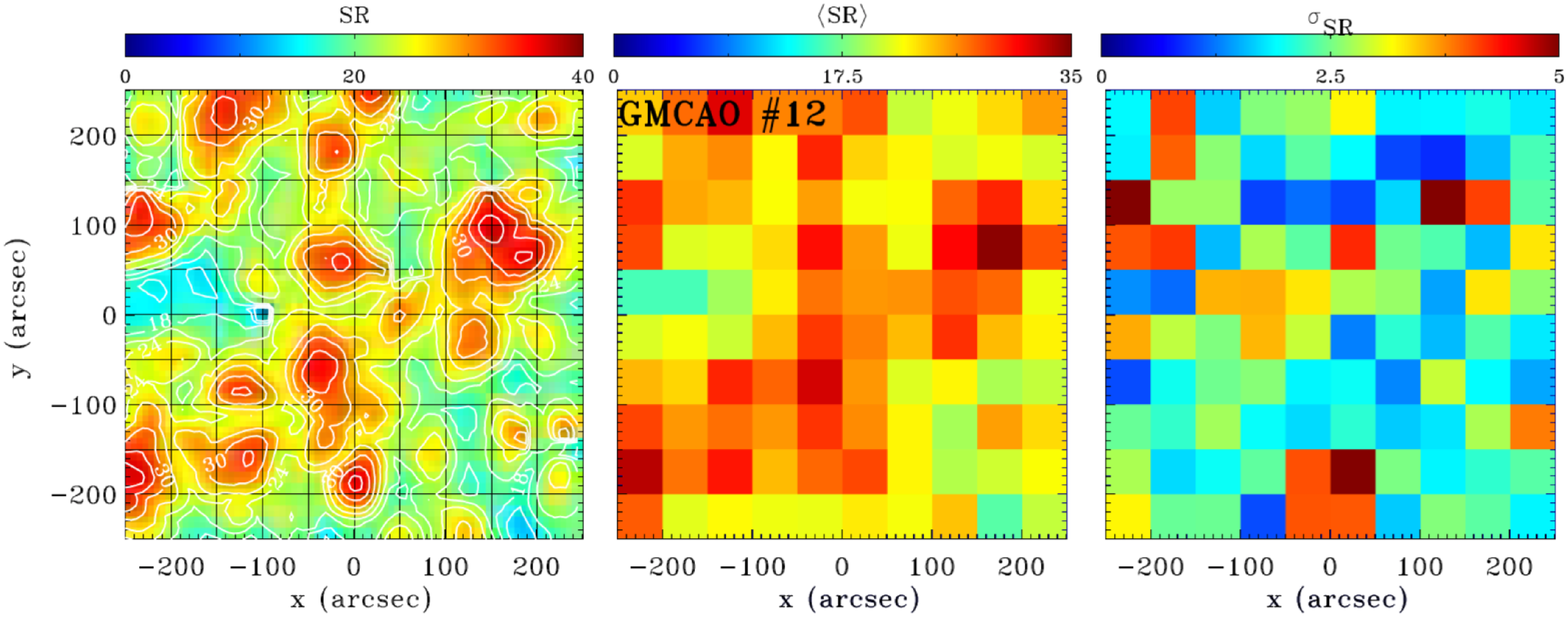}
\end{tabular}
\end{center}
\caption[example] 
{\label{fig:sr}K-band SR map and contours obtained with GIUSTO (left); average SR values (center) and standard deviation (right) measured within the 100 sub-fields. Color-bars indicate the corresponding panel scales. Only the worst (GMCAO \#4 ) and the best (GMCAO \#12) field are shown.}
\end{figure}

To evaluate the performance of the method, we built a SR map for all the fields, and performed a statistical analysis of the results: Figure~\ref{fig:sr} shows the worst (top) and the best result obtained.

\begin{table}[h]
\caption{Main inputs parameter used for GIUSTO.Column~1: Variable. Column~2: Corresponding value used in the simulations.} 
\label{param}
\begin{center}       
\begin{tabular}{|l|c|}
\hline
\rule[-1ex]{0pt}{3.5ex}Number of NGSs           & 6                   \\
\hline
\rule[-1ex]{0pt}{3.5ex}Position/Magnitude       & USNO-B1 catalog     \\
\hline
\rule[-1ex]{0pt}{3.5ex}NGSs Limit magnitude     & 18 mag (R)          \\
\hline
\rule[-1ex]{0pt}{3.5ex}Screen scale             & 0.1 m/pixel         \\
\hline
\rule[-1ex]{0pt}{3.5ex}Number of VDMs           & 6                   \\
\hline
\rule[-1ex]{0pt}{3.5ex}Sensing wavelength       & 0.5        $\mu$m       \\       
\hline
\rule[-1ex]{0pt}{3.5ex}Scientific wavelength    & 2.2        $\mu$m       \\     
\hline
\rule[-1ex]{0pt}{3.5ex}Entrance pupil diameter  & 39         $\mu$m       \\   
\hline
\rule[-1ex]{0pt}{3.5ex}Probes proximity (min)   & 10         arcsec   \\ 
\hline
\rule[-1ex]{0pt}{3.5ex}Technical FoV            & 600        arcsec   \\
\hline
\rule[-1ex]{0pt}{3.5ex}Central rejected FoV     & 120        arcsec   \\
\hline
\rule[-1ex]{0pt}{3.5ex}Scientific FoV           & 120        arcsec   \\
\hline
\rule[-1ex]{0pt}{3.5ex}Number of fields         & 100                 \\
\hline
\rule[-1ex]{0pt}{3.5ex}Number of DMs            & 3                   \\
\hline
\rule[-1ex]{0pt}{3.5ex}DMs conjugated altitudes & 0,4,12.7   km       \\
\hline
\rule[-1ex]{0pt}{3.5ex}Atmospheric profile      & 35         layers   \\
\hline
\rule[-1ex]{0pt}{3.5ex}Fried Radius             & 0.14       m        \\
\hline
\rule[-1ex]{0pt}{3.5ex}L0 max                   & 25         m        \\
\hline
\rule[-1ex]{0pt}{3.5ex}l0 min                   & 0.001      m        \\
\hline
\rule[-1ex]{0pt}{3.5ex}Atmospheric spectrum     & von Karman          \\
\hline
\end{tabular}
\end{center}
\end{table} 

\section{Conclusions}
The SR averaged value of all the 22 fields is 16.8$\pm$2.3, consistent with our previous work [\citenum{Portaluri2017}].

\begin{figure} [h]
   \begin{center}
   \begin{tabular}{c} 
   \includegraphics[height=6.5cm]{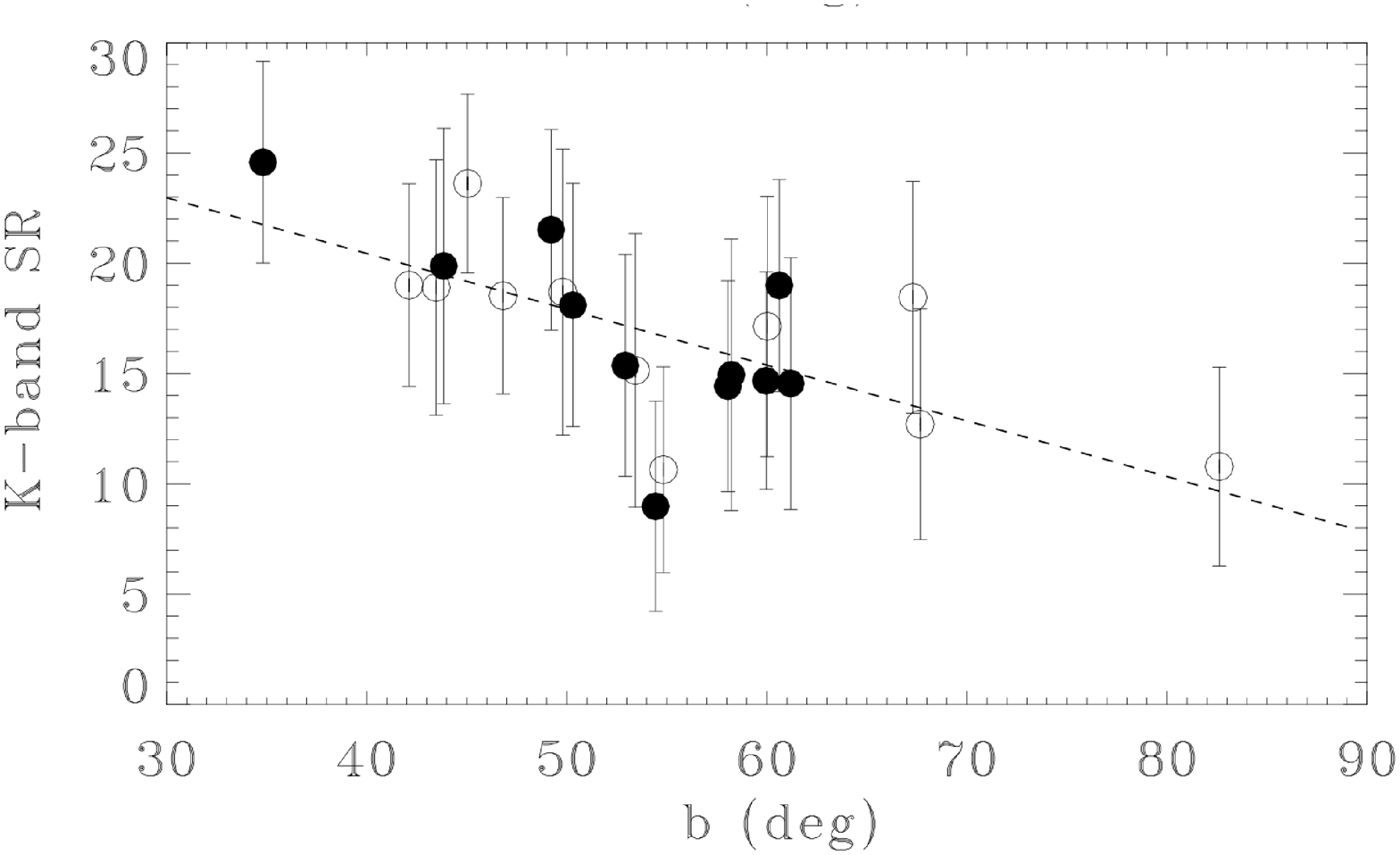}
   \end{tabular}
   \end{center}
   \caption[example] 
   { \label{fig:lat}K-band SR as a function of the galactic latitudes of the 22 GMCAO fields. Filled circles represent points with negative values of galactic latitude, while open circles are for positive values. The error bars represent $1\sigma$ value.}
   \end{figure}

The SRs anti-correlate linearly with the latitudes as expected due to the availability of stars, as clear from Figure~\ref{fig:lat}.

\begin{figure} [h]
   \begin{center}
   \begin{tabular}{c} 
   \includegraphics[height=8cm]{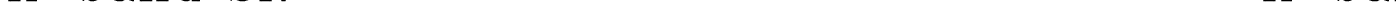}
   \end{tabular}
   \end{center}
   \caption[example] 
   { \label{fig:concl} NGS K -band magnitude (left) and radial distance from the center (right) as a function of the field SR for all the 22 GMCAO surveys. The red points were obtained with a Monte Carlo simulation using a 40-layers profile atmosphere. The solid lines represent the best-fit, while the dashed one shows the lower limit for finding NGSs in the technical field.}
   \end{figure}

   The best performance are obtained with stars that are relatively close to the scientific FoV (Figure~\ref{fig:concl}): stellar magnitudes of NGSs do not impact the performance, while mean off-axis position of NGSs plays a key role. In such a figure is also shown a comparison with a simulation made with a different atmospheric profile (the same used in [\citenum{Portaluri2017}]) to show the importance of choosing realistic parameters for the simulations in order to obtain consistent results.

The performance that can be obtained with the GMCAO method have a robust statistical confirmation: it can be applied to the next-generation of ELTs to carry out deep field observations to shed light in the puzzle of formation and evolution of galaxies.

A simple consideration is that a generic MCAO system could not use all the stars selected for AO correction, because its typical Technical FoV is of about 2 arcmin, while the 57\% of the NGS we used are further away, so we can conclude that the GMCAO method can be used in star-poor regions more that other techniques, guaranteeing an uniform correction of the field.

\bibliography{main} 
\bibliographystyle{spiebib} 

\end{document}